    \documentstyle[proceedings,numreferences]{maxent95}  

\begin{opening}
\title{ROLE AND MEANING OF SUBJECTIVE PROBABILITY\protect\\
       SOME COMMENTS ON COMMON MISCONCEPTIONS}

\author{G. D'AGOSTINI}
\institute{Dipartimento di Fisica dell'Universit\`a ``La Sapienza''\\
           Piazzale Aldo Moro 2, I-00185 Roma 
           (Italy)\footnote{Email: giulio.dagostini@roma1.infn.it.\  
  URL: http://www-zeus.roma1.infn.it/$^\sim$agostini.
\\ Contribution at the XX  International Workshop on Bayesian 
Inference and Maximum Entropy Methods in Science and Engineering, 
Gif sur Yvette, France, July 8--13, 2000.
}}

\end{opening}

\runningtitle{COMMENTS ON SUBJECTIVE PROBABILITY}

\begin{document}

\begin{abstract}
Criticisms of so called `subjective probability' come
on the one hand 
from those who maintain that probability in physics has
only a frequentistic interpretation, and, on the other, 
from those who tend to `objectivise' Bayesian theory,
arguing, e.g., that subjective probabilities are indeed based 
`only on private introspection'. Some of the common
misconceptions on subjective probability will be commented 
upon in support of the thesis that coherence is the most
crucial, universal and `objective' way to assess our confidence 
on events of any kind. 

\end{abstract}


\section{Introduction}
The role of scientists is, generally speaking, to understand 
Nature, in order to forecast as yet unobserved (`future') events,
independently of whether or not these events can be influenced.
In laboratory experiments and all technological
applications, observations
depend on our intentional manipulation of the external world.
However, other scientific activities, like 
astrophysics, are only observational. Nevertheless, to claim
that cosmology, climatology or geophysics are not Science,
because ``experiments cannot be repeated'' 
- as pedantic interpreters of Galileo's scientific method do - 
is, in my opinion, short-sighted 
(for a recent defence of this strict Galilean 
point of view, advocating `consequently' 
frequentistic methods,
see Ref.~\cite{Giunti}). The link between past observations
and future observations is provided by theory (or model). 

It is accepted that quantitative 
(and, often, also qualitative) forecasting of future observations 
is invariably uncertain, from the moment that we define 
sufficiently precisely the details of the future events. 
The uncertainty may arise because we are not 
certain about the parameters
of the theory (or of the theory itself), and/or
 about the initial state
and boundary conditions of the phenomenon we want to describe. 
But it may also be due to the stochastic nature of the theory itself, 
which would produce uncertain predictions even if all parameters and 
boundary conditions {\it were} precisely known. 
Nevertheless, the constant state of uncertainty 
does not prevent us from doing
science. As Feynman wrote, ``it is scientific only to say
what is more likely and what is less likely''.\cite{Feynman}
This observation holds not only for observations, 
but also for the values of physical quantities
(i.e. parameters of the theory which have effect on the 
real observations).    
And indeed, physicists find probabilistic statements
about, for example,  top quark mass or gravitational 
constant very natural,\cite{Maxent98}
and several equivalent expressions are currently used, such as
``to be more or less {\it confident}'', 
``to consider something more or
less {\it probable}, or more or less {\it likely}'', 
``to {\it believe} more or less something''. 
However, the subjective definition of probability, 
the only one consistent with the above expressions, is usually 
rejected because of educational bias according to which 
``the only scientific definition of probability is the 
frequentistic one,'' ``quantum mechanics only allows the 
frequency based definition of probability,'' ``probability 
is an objective property of the physical world,'' etc. 
In this paper I will comment on these and other objections
against the so called `subjective Bayesian' point of view. 
Indeed, some criticisms come from `objective Bayesians',
who have been, traditionally, in a clear majority 
during this workshop series. 

I don't expect to solve these debates in 
this short contribution, especially considering 
that many aspects of the debate  are of a psychological 
and sociological nature.  Neither will I be
able to analyse in detail every objection or to cite all the  
counter-arguments. I prefer, therefore, to focus here 
only on a few  points, referring to other papers~\cite{YR,anxiety,ajp} 
and references therein for points 
already discussed elsewhere. 

\section{Subjective Probability and Role of Coherence}
The main aim of subjective probability is to recover
the intuitive concept of probability as degree of belief. 
Probability is then related to uncertainty and not (only)
to the outcomes of repeated experiments. Since uncertainty
is related to knowledge, probability is only meaningful
as long as there are human beings interested in knowing 
(or forecasting)
something,
no matter if ``the events considered are 
in some sense {\it determined},
or known by other people.''\cite{deFinetti} 
Since - fortunately! - we do not share identical 
 states of information,
we are in different conditions of uncertainty. Probability is 
therefore only and always conditional probability, 
and depends on the different subjects interested in it 
(and hence the name {\it subjective}). This 
point of view about probability 
is not related to a single evaluation rule. In particular, 
symmetry arguments and past frequencies, as well as their combination
properly weighted by means of Bayes' theorem, can be used. 

Since beliefs can be expressed in terms of betting odds, 
as is well known and done in practice, 
betting odds 
can be seen as the most general way of making 
relative beliefs explicit, independently of 
the kind of events one is dealing with, or 
of the method used to define the odds. For example,
everybody understands Laplace's statement
concerning Saturn's mass, that ``it is a bet of 
10000 to 1 that the error of this result is not 1/100th 
of its value.''\cite{Sivia} 
I wish all experimental results to be provided in these terms,
instead of the misleading~\cite{CLWdag}
 ``such and such percent CL's.'' 
What matters 
is that the bet must be reversible
and that no bet can be arranged in such a way that 
one wins or loses with certainty. The second condition
is a general condition concerning bets. 
The first condition
forces the subject to assess the odds consistently
with his/her beliefs and also to accept the second condition:
  once he/she has fixed the odds, 
he/she must be ready to bet in either direction. 
Coherence has two important roles: the first is, so to speak,
moral, and forces people to be honest; the second is
formal, allowing the basic rules of probability
to be derived as theorems,
including the formula relating conditional probability
to probability of the conditionand and their joint probability
(note that, consistent with the use of probability 
in practice and with
the fact that in a theory where only conditional probabilities matter,
it makes no sense to have a formula that 
{\it defines} conditional probability,
see e.g. Section 8.3 of Ref.~\cite{YR} for further comments
and examples). 

Once coherence is included in the subjective Bayesian 
theory, it becomes evident that `subjective' cannot be confused
with `arbitrary', since all ingredients 
for assessing probability must be taken into
account, including the knowledge that somebody else might
assess different odds for the same events.
Indeed, 
the coherent subjectivist is far more responsible 
(and more ``objective'', in the sense 
that ordinary parlance gives to this word) 
than those who blindly use
standard `objective' methods 
(see examples in Ref.~\cite{YR}). Another source 
of objections is the confusion between `belief' and `imagination', 
for which I refer to Ref.~\cite{anxiety}. 

\section{Subjective Probability, Objective Probability, Physical
Probability}
To those who insist on objective probabilities I like 
to pose practical questions, such as how they 
would evaluate probability
in specific cases, instead of letting  them pursue
mathematical games. 
Then it becomes clear that, at most, probability evaluations
can be intersubjective, if we all share the same education 
and the same real or conventional state of information.
The probability that a molecule of $N_2$ at a certain temperature
has a velocity in a certain range {\it seems} objective: 
take the Maxwell velocity p.d.f., make an integral 
and get a number, say $p=0.23184\ldots$ 
This mathematical game gets immediately
complicated if one thinks about a real vessel, containing real gas,
and the molecule velocity measured in a real experiment. 
The precise `objective' number obtained from the above integral
might no longer correspond to our confidence that the 
velocity is really in that interval. The idealized ``physical
probability'' $p$ can easily be a misleading ``metaphysical'' 
concept which does not correspond to the confidence of 
real situations. In most cases, in fact, $p$ is a number that 
one gets from a model, or a free parameter of a model. 
Calling $E$ the {\it real} event and $P(E)$ the probability 
we attribute to it, the idealized situation corresponds to 
the following conditional probability:
\begin{equation}
P(E\,|\,\mbox{Model}\rightarrow p) = p\,.
\end{equation}
But, indeed, our confidence on $E$ relies on our confidence
on the model:
\begin{equation}
P(E\,|\,I) = 
\sum_{\mbox{Models}}P(E\,|\,I,\mbox{Model}\rightarrow p)\cdot
P(\mbox{Model}\rightarrow p\,|\,I)  \,,
\label{eq:sumModels}
\end{equation}
where $I$ stands for a background state of information 
which is usually implicit in all probability assessments. 
Describing our uncertainty on the {\it parameter} $p$
by a p.d.f. $f(p)$ (continuity is assumed for simplicity), 
the above formula can be turned into
\begin{equation}
P(E\,|\,I)=\int_0^1\!P(E\,|\,p,I)\cdot f(p\,|\,I)\,\mbox{d}p\,.
\label{eq:intp}
\end{equation}
The results of Eqs. (\ref{eq:sumModels}) and (\ref{eq:intp}) 
really express the meaning of probability, describing our beliefs,
and upon which (virtual) bets can be set (`virtual' because it is
well known that real bets are delicate decision problems
of which beliefs are only one of the ingredients).  

For those who still insist that probability is a property of 
the world, I like to give the following example, 
readapted from Ref.~\cite{Scozzafava}. Six 
externally indistinguishable boxes each contain 
five balls, but with
differing numbers of black and white balls
(see Ref.~\cite{ajp}
for details and for a short introduction of Bayesian inference based
on this example). One box is chosen at random. What will be its
white ball  content? If we extract a ball, what is the probability
that it will be white? 
Then a ball is extracted and turns out to be white. 
The ball is reintroduced into 
the box, and the above two questions are asked again. 
As a simple application of Bayesian inference, the probability 
of extracting a white ball in the second extraction becomes 
$P(E_2=W)=73\%$, 
while it {\it was} $P(E_1=W)=50\%$ for the first extraction. 
One does not need to be a Bayesian to solve this simple
text book example, and everybody will agree on the two values 
of probability (we have got ``objective'' results, so to say). 
But it is easy to realize that these probabilities do not 
represent a `physical property of the box', but 
rather a `state of our mind', which changes as the 
extractions proceed. In particular, `measuring', or `verifying', 
that $P(E_2=W)=73\%$ using the relative frequency makes no sense.  
We could imagine a large number of extractions. 
It is easy to understand, given our prior knowledge
of the box contents, that the relative frequencies ``will 
tend'' (in a probabilistic sense) to $\approx 20\%$, 
$\approx 40\%$,   $\approx 60\%$,  $\approx 80\%$, or
  $\approx 100\%$, but `never' 73\%~\footnote{Does this violate 
Bernoulli's theorem? I leave the solution to this apparent 
paradox as amusing problem to the reader.}. 
This certainly appears to be a paradox 
to those who agree 
 that   $P(E_2=W)=73\%$ is
the `correct' probability, but still 
maintain that probability as degree of belief is a useless concept. 
In this simple case the six a priori probabilities
$p_i=i/5$, with $i=0,1,\ldots,5$, can be seen as the 
possible ``physical probabilities'', but the ``real'' probability
which determines our confidence on the outcome is given by
a discretized version of Eq.~(\ref{eq:intp}), 
with $f(p_i)$ changing from one extraction to the next. 

I imagine that at this point some readers might react 
by saying that the above example proves that only the frequentistic 
definition of probability is sensible, because the relative 
frequency will tend for $n\rightarrow \infty$ to the `physical  
probability', identified in this case by the 
white ball ratio in the box. 
But this reaction is quite na\"\i ve. First, a definition
valid for $n\rightarrow\infty$ is of little use for practical 
applications (``In the long run we are all dead''\cite{Keynes}).
Second, it is easy to show~\cite{ajp} that, 
for the cases in which the `physical probability' can be checked
and the number of extraction is finite, though large,
the convergence behaviour of the frequency based evaluation 
is far poorer than the Bayesian solution and can also be
in paradoxical contradiction with the 
available status of information. 
Moreover, only the Bayesian theory answers consistently,
in unambiguous probabilistic terms, the 
legitimate question ``what is the box content?'',
since the very concept of probability of hypotheses  
is banished in the frequentistic approach. 
Similarly, only in the Bayesian approach does it make sense 
to express in a logical, consistent way the confidence 
on the different causes of observed events and on the 
possible values of physics quantities (which are unobserved entities).
To state that ``in high energy physics, where experiments are
repeatable (at least in principle) the definition of probability 
normally used''\cite{PDG} is to ignore the fact that 
the purpose of experiments is not to predict 
which electronic signal will
come out next from the detector (following the analogy of 
the six box example), but rather to narrow the range in which 
we have high confidence that the physics quantities 
lie.\footnote{To be more rigorous, simple laboratory experiments 
can be performed in conditions of repeatability\cite{DIN-ISO}, 
but thinking of repeating very complex particle physics experiments
run for a decade make no sense (even in principle!).
Perhaps the remark ``in principle'' in the above quotation 
from Ref.~\cite{PDG} is to justify Monte Carlo simulation
of the experiments. But one has to be aware that a Monte
Carlo program is nothing but a collection of our best beliefs
about the behaviour of the studied reaction, background reactions
and apparatus.}
  
\section{Observed Frequencies, Expected Frequencies, Frequentistic
Approach and Quantum Mechanics}
Many scientists think they are frequentists because they are 
used to assessing their beliefs in terms of expected frequencies,
without being aware of the implications 
for a sane person of sticking strictly to frequentistic ideas. 
Certainly, past frequencies can be a part of the information
upon which probabilities can be assessed~\cite{ajp,CLWdag}.
Similarly, probability theory teaches us how to 
predict future 
frequencies from the assessment of beliefs, under well defined
conditions. But identifying probability with frequency is like
confusing a table with the English word `table'. 
This confusion leads some authors, because they lack 
other arguments to save the manifestly sinking boat of the 
frequentistic collection of {\it adhoc-eries}, 
to argue\footnote{A similar  desperate attempt is  
try to throw a bad shadow over Bayesian theory, 
saying that in this theory ``frequency and probability
are completely disconnected''\cite{James}, using as argument 
an ambiguous sentence picked up from the large Bayesian 
literature, and severed from its context.}
that ``probability in quantum mechanics is frequentistic 
probability, and is defined as long-term frequency. Bayesians
will have to explain how they handle that problem,
and they are warned in advance.''\cite{James}

Probability deals with the belief that an event may happen, 
given a particular state of information. 
It does not matter if the fundamental laws are 
`intrinsically probabilistic', or if it is just a limitation 
of our present ignorance. The impact on our minds
remains the same.
If we think of two possible events resulting 
from a quantum mechanics experiment, and say
 (after computing no matter 
how complicated calculations to also take into
account detector effects) that $P(E_1)\gg P(E_2)$, 
this simply means that we feel more 
{\it confident} in $E_1$ than in $E_2$, 
or that we will be {surprised} if $E_2$ happens instead of $E_1$.
If we have the opportunity to repeat the experiment we 
believe that events of the `class' $E_1$ will happen more frequently 
than events of class $E_2$.  
This is what we find carefully reviewing the relevant 
literature and discussing with theorists: probability is 
`probability', although it might be expressed in terms of expected 
frequencies, as discussed above. 
Take for example Hawking's 
{\sl A Brief History of Time},\cite{Hawking}
(which a statistician\cite{Berry} said should be called 
`a brief history of beliefs', so frequently 
do the words belief, believe and synonyms appear in it).
For example:
``In general, quantum mechanics 
does not predict a single definite result for an observation. 
Instead, it predicts a number of different possible 
outcomes and tells us how likely each of these is''~\cite{Hawking}.
Looking further  to the past, 
it is worth noting  the concept of ``degree of truth''
introduced by von Weizs\"acker, as reported by 
Heisenberg~\cite{Heisenberg}. It is difficult to find any difference 
between this concept and the usual degree of belief, especially 
because both Heisenberg and von Weizs\"acker were fully aware
that ``nature is earlier than man, but man is earlier than 
natural science''~\cite{Heisenberg}, in the sense that 
science is done by our brains, mediated by our senses.  
It is true that, reading some 
text books on quantum mechanics one gets the idea
that ``probability is frequentistic probability''\cite{James},
but one should remember the remarks at the beginning 
of this section, and the fact that many authors 
have used, uncritically, the dominant ideas on probability
in the past decades. But some authors also try to account 
for probability of {\it single} events, instead of `repeated 
events', and have to admit that this is possible 
if probability is meant as degree of belief. 

In conclusion, invoking intriguing fundamental aspects
of quantum mechanics in the discussion of  inferential frameworks
shows little awareness of 
the real issues involved in the two classes of problems. 
First, the debate about the interpretations of quantum mechanics
is far from being settled~\cite{tHooft}.
Second, as far  as `natural science' is concerned, 
it doesn't really matter if nature is deeply deterministic or 
probabilistic, as eloquently said by Hume: 
``Though there is no such thing as Chance in the world; 
our ignorance of the real cause
of any event has the same influence on the understanding, 
and begets a like species of belief or opinion.''\cite{Hume}

\section{Who is Afraid of Subjective Bayesian Theory?}
``It is curious that, even when different workers 
are in substantially
complete agreement on what calculations should be done, 
they may have radically different views as to what we are actually 
doing and why we are doing it.''~\cite{Jaynes} 
It is indeed 
surprising that strong criticisms of subjective probability
come from people who essentially agree that probability
represents ``our degree of confidence''~\cite{Jeffreys}
and that Bayes' theorem is the proper inferential tool. 
I would like to comment here on criticisms 
(and invitations to convert\ldots) which 
I have received from objective Bayesian
friends and colleagues, and which can be traced
back essentially to the same source.~\cite{Jaynes}
The main issue in the debate is the choice of the {\it prior}
to enter in the Bayesian inference. I prefer
subjective priors because they seem to me to correspond more
closely to the spirit of the Bayesian theory and 
the results 
of the methods based on them are more 
reliable and never paradoxical~\cite{anxiety}. 
Nevertheless,  I agree, in principle,
that a ``concept of a `minimal informative' prior specification
 -- appropriately defined!''\cite{BS} can be useful in 
particular applications.
The problem is that those
who are not fully aware of the intentions and limits of 
the so called {\it reference priors} tend to perceive
the Bayesian approach as dogmatic. 
Let us analyse, then, some of the criticisms. 
\begin{itemize}
\item[-]
``Subjective Bayesians have settled into a position intermediate
between orthodox statistics and the theory 
expounded here.''\cite{Jaynes}
I think exactly the opposite. Now, it is obvious
that frequentistic methods are conceptually a mess, a collection
of arbitrary prescriptions. But those who stick too strictly to the 
theory expounded in Ref.~\cite{Jaynes} tend to give up the 
real (unavoidably subjective!) knowledge of the problem
in favour of mathematical convenience or  
blindly following the stance taken by the leading
figures in their school of thought.
This is exactly what happens
 with practitioners using blessed 
`objective' frequentistic `procedures' (for example, see
Ref.~\cite{anxiety} for a discussion on the misuse
of Jeffreys' priors). 
\item[-]
``While perceiving that probabilities cannot represent 
only frequencies, they [subjective Bayesians] still regard 
sampling probabilities as representing frequencies of 
`random variables'''.\cite{Jaynes} The name`random
variable' is avoided by the most 
authoritative subjective Bayesians~\cite{deFinetti}
and the terms `uncertain (aleatoric) numbers' 
and `aleatoric vectors' (form multi-dimensional cases)
are currently used.
Even the idea of `repeated events' is rejected~\cite{deFinetti},
as every event is unique, though one might think of 
classes of analogous events to which we can attribute the same
{\it conditional} probability, but these events are usually 
stochastically dependent (like the outcomes
black and white in the six box example of Ref.~\cite{ajp}).
In this way the ideas of uncertainty
and probability are completely disconnected from 
that of {\it randomness} \`a la von Mises.~\cite{vonMises}   
Nevertheless, I admit that there are authors, including
myself~\cite{YR}, who mix the terms `uncertain numbers' and 
`random variables', to make life easier for those who are not
accustomed to the concept of uncertain numbers, since the
formal properties (like p.d.f., expected value, 
variance, etc) are the same for the two objects. 
\item[-] 
``Subjective Bayesians face an awkward ambiguity at the beginning
of a problem, when one assigns prior probabilities. If these 
represent merely prior opinions, then they are basically 
arbitrary and undefined''. \cite{Jaynes} Here the
confusion between subjective and arbitrary, discussed above,
is obvious.
\item[-]
``It seems that only private introspection could assign them
and different people will make different assignments''.
No knowledge, 
no science, and therefore no probability, is conceivable 
if there is no brain to analyse the {\it external} world. 
Fortunately there are no two identical brains (yet), 
and therefore no two identical states of knowledge are conceivable, 
though intersubjectivity can be achieved in many cases. 
\item[-]
``Our goal is that inferences are to be completely `objective' 
in the sense that two persons with the prior information 
must assign the same prior probability.''~\cite{Jaynes} 
This is a very na\"\i ve idealistic statement of little
practical relevance. 
\item[-]
``The natural starting point in translating a number of pieces 
of prior information is the state of 
complete ignorance.''~\cite{Jaynes}
When should we define the state of complete ignorance? 
At conception or at birth? How much is learned 
and how much was already coded in the DNA? 
\end{itemize} 

\section{Conclusions}
To conclude, I think that none of the above criticisms is really
justified. As for criticisms put forward by frequentists
or by self-designed frequentistic practitioners (who
are more Bayesian than they think they are~\cite{Maxent98}) 
there is little
more to comment in the context of this workshop. 
I am much more interested in making some final comments 
addressed to fellow Bayesians who do not share 
some of the ideas expounded here. I think 
that users and promoters of Bayesian methods
of the different schools should make an effort to smooth
the tones of the debate, because the points we have in common 
are without doubt many more, and more relevant, than those on
which there is disagreement. Working on similar problems and 
exchanging ideas will certainly help us to understand each other. 
There is no denying that Maximum Entropy methods are very useful 
in solving many complicated practical problems, as 
this successful series of workshops has demonstrated. 
But I don't see any real contradiction with coherence: 
I am ready to take seriously the result of any method,
if the person responsible for the result
is honest and is ready to make any combination
of reversible bets based on the declared probabilities. 



\begin{thebibliography}{99}

\bibitem{Giunti} 
C. Giunti, Proc. Workshop on Confidence Limits,
CERN, Geneva, Switzerland, January 2000. CERN Report 2000--005,
May 2000, pp. 63--72, e-print arXiv: hep-ex/0002042.
%
\bibitem{Feynman}
R. Feynman, {\sl The character of physical law}, MIT Press, 1967.
%
\bibitem{Maxent98}
G. D'Agostini, Proc. of the
XVIII International Workshop on Maximum Entropy and Bayesian
Methods, Garching (Germany), July 1998. 
(Kluwer Academic Publishers, Dordrecht, 1999), pp. 157--170,
e-print arXiv: physics/9811046.
%
\bibitem{YR}
G. D'Agostini, CERN Report 99--03, July 1999, electronic
version available at the author's URL.
%
\bibitem{anxiety}
G. D'Agostini, Rev. R. Acad. Cienc. Exact. Fis. Nat., Vol. 93, 
Nr. 3,  1999, pp. 311--319, e-print arXiv: physics/9906048.
%
\bibitem{ajp}
G. D'Agostini, Am. J. Phys. {\bf 67} (1999) 1260-1268,
e-print arXiv: physics/9908014.
%
\bibitem{deFinetti} 
B. de Finetti, {\sl Theory of Probability} 
(J. Wiley \& Sons, 1974).~\label{bib:deFinetti}
%
\bibitem{Sivia}
D.S. Sivia, {\sl Data analysis -- a Bayesian tutorial},
Oxford University Press, 1997.
%
\bibitem{CLWdag} 
G. D'Agostini, Proc. Workshop on Confidence Limits,
CERN, Geneva, Switzerland, January 2000. CERN Report 2000--005,
May 2000, pp. 3--23.
%
\bibitem{Scozzafava}
R. Scozzafava, Pure Math. and Appl., Series C 2, 1991, pp. 223-235.
%
\bibitem{Keynes}
J.M. Keynes, {\sl A tract on Monetary Reform} (Macmillan, London, 1923).
%
\bibitem{PDG}
Particle Data Group, D.E. Groom et al., 
{\it Eur. Phys. J.} {\bf C15} (2000) 1 (Section 28).
%
\bibitem{DIN-ISO}
DIN Deutsches Institut f\"ur Normung, 
{\it ``Grundbegriffe der Messtechnick -- Behandlung von
Unsicheratine bei der Auswertung von Messungen''},
(DIN 1319 Teile 1--4), Beuth Verlag GmBH, Berlin, Germany, 1985.\\
ISO International Organization for Standardization, 
{\it International vocabulary of basic and genaral terms in 
metrology},'' Geneva, Switzerland, 1993.
%
\bibitem{James}
F. James,   Proc. Workshop on Confidence Limits,
CERN, Geneva, Switzerland, January 2000. CERN Report 2000--005,
May 2000, pp. 1--2.

\bibitem{Hawking}
S.W. Hawking, {\sl A Brief history of time}, Bantam Doubleday Dell, 1988.

\bibitem{Berry}
D.A. Berry, Talk at the Annual Meeting of American 
Statistical Association, Chicago, August 1996.

\bibitem{Heisenberg}
W. Heisenberg, {\sl Physics and philosophy. The revolution
in modern science}, 1958 (Harper \& Row Publishers, New York, 1962). 
%
\bibitem{tHooft}
See e.g. G. 't Hooft, {\sl In search of the ultimate building blocks}, 1992
(Cambridge University Press, 1996).
%
\bibitem{Hume}
D. Hume, {\sl Enquiry concerning human understanding}'', 1748. 
%

\bibitem{Jaynes}
E.T. Jaynes, {\sl Probability Theory: The logic of Science},
Chapter 12 on {\it Ignorance priors and
transformation groups}, 
(book available at http://bayes.wustl.edu/etj/prob.html and 
http://omega.albany.edu:8008/JaynesBook, but Chapter 12
is missing since beginning of 2000). 
%
\bibitem{Jeffreys}
H. Jeffreys, {\sl Theory of probability}, 1939 
(Clarendon Press, Oxford, 1998).
%
%
\bibitem{BS}
J.M. Bernardo and A.F.M. Smith, {\sl Bayesian Theory},
John Wiley \& Sons, 1994. 
%
\bibitem{vonMises}
R. von Mises, {\sl Probability, Statistics and Truth}, 1928,
(George Allen \& Unwin, 1957), second edition.
%
\end{thebibliography}
\end{document}